%% file: main-expanded.tex
\documentclass[a4paper]{article}

\usepackage[margin=1.18in]{geometry} 

\usepackage{booktabs}
\usepackage[T1]{fontenc}
\usepackage{lmodern}
\usepackage{hyperref}
\hypersetup{
    unicode,
    hidelinks,
    colorlinks,
urlcolor=blue, 
    linkcolor=blue,
    citecolor=blue,
    pdfauthor={Marco L. Della Vedova, Mattias Wahde},
    pdftitle={Naturalness Indicators of Forests in Southern Sweden derived from the Canopy Height Model},
    pdfkeywords={machine learning, forests, interpretability, canopy height model, remote sensing},
}

\usepackage[numbers,square]{natbib}
\usepackage{graphicx, color}
\usepackage{url}
\usepackage{amsmath,amsthm,amssymb}
\usepackage{tabularx}
\usepackage{siunitx}
\usepackage{import}
\usepackage{svg}
\usepackage{pgfplots}
\pgfplotsset{compat=1.18}

\title{\textbf{Naturalness Indicators of Forests in Southern Sweden derived from the Canopy Height Model}}

\author{
    Marco~L.~Della~Vedova$^{1}$\thanks{Corresponding author: marco.dellavedova@chalmers.se} 
    \and Mattias Wahde$^1$
}

\date{
    $^1$Chalmers University of Technology \\ 
    Gothenburg, Sweden\\[2ex]\today
}

\begin{document}
\maketitle

\begin{abstract}
Forest canopies embody a dynamic set of ecological factors, acting as a pivotal interface between the Earth and its atmosphere. 
They are not only the result of an ecosystem’s ability to maintain its inherent ecological processes, structures, and functions but also a reflection of human disturbance. 
This study introduces a methodology for extracting a comprehensive and human-interpretable set of features from the Canopy Height Model (CHM), which are then analyzed to identify reliable indicators for the degree of naturalness of forests in Southern Sweden.
Utilizing these features, machine learning models -- specifically, the perceptron, logistic regression, and decision trees -- are applied to predict forest naturalness with an accuracy spanning from 89\% to 95\%, depending on the area of the region of interest.
The predictions of the proposed method are easy to interpret, something that various stakeholders may find valuable.
\\[2ex]\noindent\textbf{Keywords:} machine learning, forests, interpretability, canopy height model, remote sensing.
\end{abstract}

\section{Introduction}
\label{sec:intro}

\begin{figure}[t]
    \centering
    {\small
    \includegraphics[width=\linewidth]{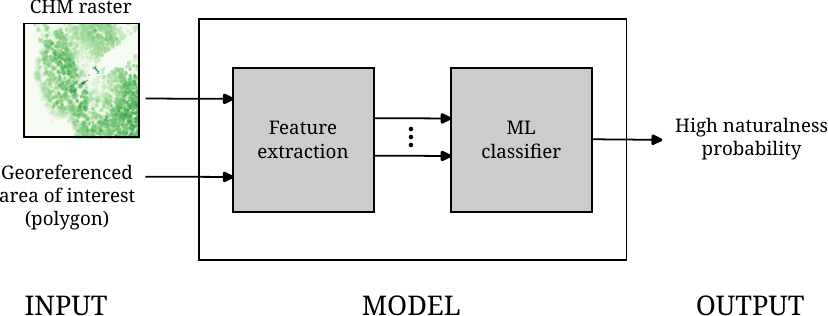}
    }
    \caption{Overview of the proposed method. The classifier is trained in a supervised machine learning fashion.}
    \label{fig:overview}
\end{figure}

Forests, as complex and dynamic ecosystems, play a pivotal role in maintaining ecological balance, biodiversity, and global climate regulation. 
In the context of forest ecosystems, \textit{naturalness} refers to the degree to which an ecosystem retains its inherent ecological processes, structures, and functions in the absence of significant human intervention~\citep{mcrobertsAssessingForestNaturalness2012}.
The assessment of the naturalness of forests has emerged 
as an important endeavor both for defining nature protection areas, considering escalating anthropogenic pressures and loss of biodiversity, and in terms of forest management by national forest agencies worldwide.

Traditionally, the assessment of forest naturalness is carried out by experts with field inventories that are collected in national forest inventories. Various methods have been suggested for evaluating naturalness, yet there is ongoing debate surrounding the merits and drawbacks of these various approaches.
Nevertheless, the typical traits of naturalness are considered to be: diversity in trees (species composition, stand structure), presence of dead wood, landscape age structure, and wildlife (fauna) composition~\citep{barretteNaturalnessAssessmentPerformed2020}.
Very similar traits have been used in the literature to define the concepts of \textit{ecological integrity} of forests and \textit{forest integrity}~\citep{fregoBryophytesPotentialIndicators2007,tierneyMonitoringEvaluatingEcological2009}, which are closely related to the definition of naturalness used in this paper.
In any case, assessing these characteristics traditionally requires field work, which is costly and time-consuming.

On the other hand, more and more digital data about lands and their usage are collected by remote sensing with satellites and aircraft, both manned and UAVs~\citep{lechnerApplicationsRemoteSensing2020}. 
There is therefore an opportunity to apply techniques from artificial intelligence (AI) to assess forest naturalness with automated and semi-automated methods. 
In fact, these techniques have been extensively used for similar processes~\citep{zhangArtificialIntelligenceRemote2022}, such as land use classification~\citep{zhangObjectbasedConvolutionalNeural2018}, 
leaf-area index estimation~\citep{chenSpatiotemporalPredictionLeaf2015},  
flooding prediction~\citep{saraviUseArtificialIntelligence2019, zhangInterpretableDeepSemantic2022a}, etc.
Specifically regarding forests, AI methods have been proposed for stand delineation~\citep{olofssonForestStandDelineation2014} and segmentation~\citep{mustonenAutomaticSegmentationForest2008,dechesneForestStandSegmentation2016},
prediction of tree species richness~\citep{brugereImprovedPredictionTree2023},
drone applications~\citep{bucheltExploringArtificialIntelligence2024}, 
and biomass estimation~\citep{hongCombiningMultisourceData2023}.
However, interestingly, not many studies address the problem of automatic assessment of forest naturalness.
The assessment of ecological function levels, which is related to naturalness, has been proposed by~\citep{fangAssessmentForestEcological2023}, where an overall accuracy of 0.82 was obtained using a random forest classifier (an ensemble of decision trees) applied to a multi-source dataset.

This paper introduces a novel automated method to assess forest naturalness from a set of features extracted from Canopy height model (CHM) with \SI{1}{m} resolution as single source of data. 
The CHM is a georeferenced raster in which each pixel represents the height of the trees' canopy relative to the ground. 
CHM data can be obtained from airborne laser scanning surveys, which are becoming increasingly common.
Note that with a \SI{1}{m} resolution CHM it is possible to distinguish individual trees, as also demonstrated by \citet{ozdemirModellingTreeSize2013}.

Figure~\ref{fig:overview} shows an overview of the proposed method: It takes as input the CHM and the area of interest, defined by a georeferenced polygon representing the boundary of the forest to analyze, and it produces as output the predicted probability of a forest to be of high naturalness.
Internally, the method is composed of two components: The feature extraction and the machine learning (ML) classifier.
The ML classifier is trained on forests with known high or low degree of naturalness (i.e., positive and negative examples of highly natural forests), taken from existing national inventories (see Section~\ref{sec:data}).
While the degree of naturalness would perhaps be better measured on a continuous scale, e.g., from zero to one where zero represents ``not natural at all'' and one represents ``totally natural'', in this study we consider binary classifiers since forests in the data sources are not labeled with a continuous value for their naturalness.
Nevertheless, continuous-valued degrees of naturalness can very well be derived from such classifiers, as described in Section~\ref{sec:confidenceScore}.

The main benefits of the proposed method are that 
(1) it is based on a single source of data (the CHM), thus simplifying the requirements on data acquisition and processing; 
(2) it returns a probability of high naturalness, which can be used to measure naturalness on a continuous scale; and
(3) it is interpretable, meaning that it is easy to understand its outcomes (glass box model).

There are multiple advantages with having an interpretable method.
The distinctive characteristic of such methods is that their inner workings are human-understandable. 
This is in contrast to the currently popular black box approaches based on deep neural networks (DNNs), with millions (or even billions) of non-linearly interconnected computational elements (artificial neurons), making it near-impossible for a human to follow their reasoning~\citep{Rudin2019}. 
We believe that, in the context of assessing forest naturalness, understanding the reasons \textit{why} a forest has been classified as having a certain degree of naturalness is of great value for many stakeholders, such as environmental experts, policy makers, forestry companies, and engineers that develop and maintain the classification method.
With a human-understandable method, environmental experts can identify key features of forests that are associated with naturalness. 
Policy makers can enhance transparency and accountability of forest monitoring and assessment programs, and develop more targeted conservation strategies.
The transparency of the system is important also from the perspective of forestry companies to understand why conservation strategies are in place, and why they can or cannot cut a particular forest.
Last, understanding the inner workings of the system is important for the engineers who develop, debug, maintain, and extend it.

In summary the research questions we address in this paper are the following:
(i) Is it possible to identify highly natural forests from the CHM with an interpretable method?
(ii) Which of the features of a forest, extracted from the CHM, are indicators of naturalness?

The remainder of the paper is organized as follows: In Section~\ref{sec:data} we describe the data used here, and
we also provide information about how the data set has been split into training, validation, and test sets.
Then, in Section~\ref{sec:method}, the method is described, starting with the definition of the features used,
followed by a presentation of the different machine learning models applied here. Next, in Section~\ref{sec:results}
we present the results. A discussion is given in Section~\ref{sec:discussion}, and then the conclusions in
Section~\ref{sec:conclusion}.

\begin{figure}[t]
    \centering
    \setlength{\fboxsep}{0pt}
    \fbox{\includegraphics[width=0.325\linewidth]{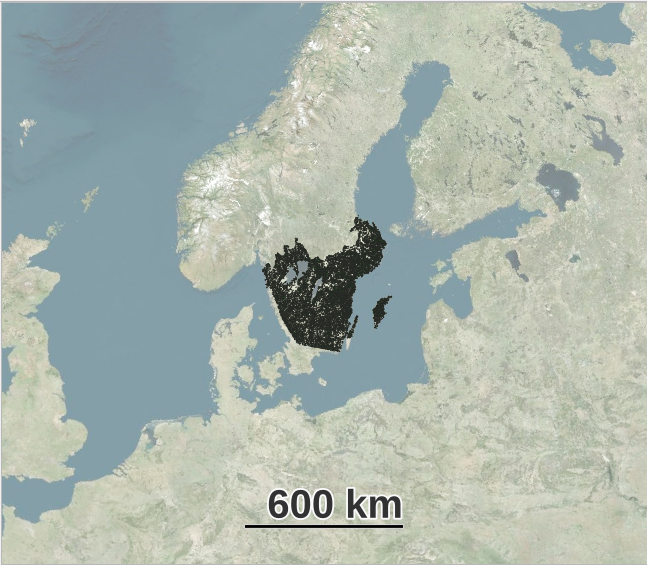}}
    \hfill
    \fbox{\includegraphics[width=0.325\linewidth]{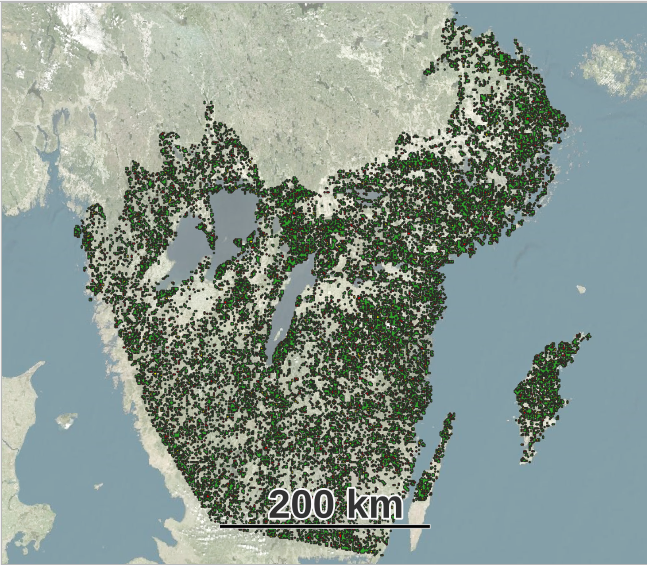}}
    \hfill
    \fbox{\includegraphics[width=0.325\linewidth]{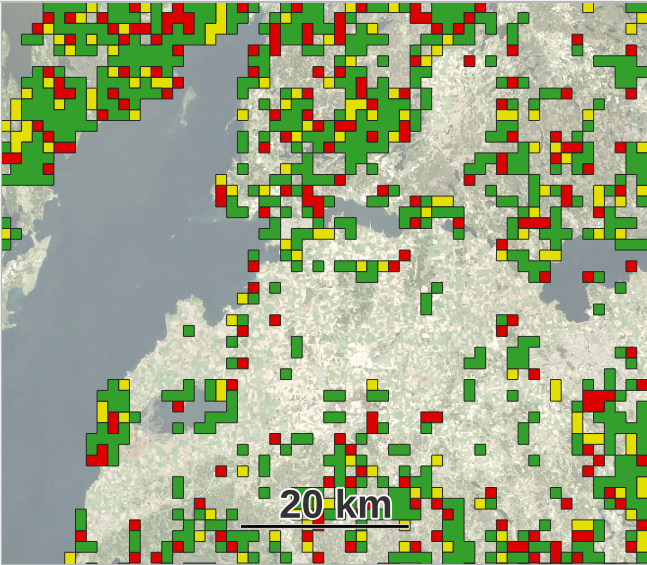}}
    \caption{The study area in Southern Sweden: Continental view (left), overview (center), and zoom (right). Training, validation, and test regions are shown in (bright) green, yellow, and red, respectively. All the three images are centered in 58.4881~N, 15.1000~E; EPGS:3006 projection.}
    \label{fig:study-area}
\end{figure}

\section{Materials}
\label{sec:data}

The study area is located in Southern Sweden 
(\ang{56.17}~$-$ \ang{60.78}~N,
\ang{10.99}~$-$ \ang{19.20}~E); see also Figure~\ref{fig:study-area}.
Forests of this region are predominantly coniferous, with Norway spruce (Picea abies) and Scots pine (Pinus sylvestris) accounting for around 75\% of the total standing tree volume. However, there is also a significant presence of broadleaved trees, particularly in southernmost Sweden. 
These include birch (Betula), European beech (Fagus sylvatica), oak (Quercus spp.), ash (Fraxinus excelsior), and alder (Alnus spp.)~\citep{drosslerTreeSpeciesMixtures2010}.

The data used in this work consists of CHM combined with
polygons that delimit the forest areas under consideration and are also associated with a class label
(high or low naturalness). Below, we first describe the CHM in Section~\ref{sec:chm} and
the polygons in Section~\ref{sec:labels-data}. The process used for dividing the data in the usual training, validation, and test splits is described in Section~\ref{sec:tr-va-te-split}.

\subsection{Canopy Height Model (CHM)}
\label{sec:chm}
The main source of data is the CHM. This data set is derived from a large set of georeferenced raster images containing information about the height of the trees.
The ground resolution is 1 meter and the measurement precision is \SI{0.1}{\meter}.
In other words, each pixel corresponds to a $1\times1$~m square on the ground and its (integer) value is the height of the upper canopy of the trees relative to the ground, in decimeters.
In addition, information about the date of the survey is available as metadata.
The data set was collected through airborne laser scanning from 2018 to 2022 and it is provided by Skogsstyrelsen, the Swedish Forest Agency, as open data\footnote{Available at \url{https://www.skogsstyrelsen.se/sjalvservice/karttjanster/skogliga-grunddata/}, \textit{tr\"adh\"ojd} layer.}.
Details about the airborne laser scanning are listed in Table \ref{tab:laserscan_details}.
Data quality of the CHM, in terms of the root mean square difference in the canopy height as measured by the laserscan and by 2,786 dedicated field survey, has been estimated to be about 8\%\footnote{Cf. \url{https://www.skogsstyrelsen.se/globalassets/om-oss/regeringsuppdrag/uppdatering-av-skogliga-grundata/kvalitetsbeskrivning-skogliga-skattningar-laserdata-20240115.pdf}}.

\begin{table}[hbt]
\caption{\label{tab:laserscan_details}
    Lidar data acquisition parameters.
}
\medskip
\centering
\small
\begin{tabular}{ll}
    \toprule
Point density & 1-2 points per square meter\\
     Flying altitude & ca. \SI{3000}{\meter} above ground\\
     Scanning angle & maximum $\pm\ang{20}$\\
     Side overlap & at least 20\%\\
     Footprint on the ground & $\leq$ \SI{0.75}{\meter} depending on flying altitude\\
     \bottomrule
\end{tabular}
\vspace{2ex}

\scriptsize{More details at \url{https://www.lantmateriet.se/globalassets/geodata/geodataprodukter/hojddata/quality_description_lidar.pdf}}
\end{table}

\subsection{Labels -- ground truth}
\label{sec:labels-data}

To collect positive and negative examples of highly natural forests, we use data from different sources.
The data contain georeferenced geometries (polygons and multi-polygons) defined by environmental experts in field campaigns. 
The coverage of these data is limited to selected regions in Sweden, due to the expensive (manual) detection process.

In detail, for positive examples, i.e., forests with high naturalness, we consider the following sources: 
(1) \textit{Naturvardsverket}, habitat-classed areas within Natura~2000, a network of protected areas throughout the EU\footnote{See \url{https://ec.europa.eu/environment/nature/natura2000/index_en.htm}}; 
(2) \textit{Storskogsbruket}, inventory of key habitats conducted by forestry companies; and 
(3) \textit{SksBorealSyd}, inventory of key habitats in the South Boreal region conducted by the Swedish Forest Agency.

Natura~2000 forests are manually annotated and include details about the inventory method employed, along with the assigned naturalness classification for the forest stand.
Among all the available forests, we limited our scope by filtering on attributes:
We considered only those annotated in the field or reviewed at the desk\footnote{Attribute \textit{Kartering} (=``mapping") $\in$ [2, 3, 4]}, which were assigned a high natural value\footnote{Attribute \textit{Naturtypss} (=``natural type") $\in$ [1, 2]}.
40\% of the polygons contain taiga (habitat code 9010) which could be both coniferous, deciduous, and mixed forest. The rest of the habitat is mostly coniferous-dominated forest.
In \textit{Storskogsbruket} approximately 14\% of the polygons are dominated by deciduous forest and the remainder is dominated by coniferous forest.
In \textit{SksBorealSyd} approximately 43\% of the polygons are predominantly deciduous forest and the rest is mainly coniferous forest.

For negative examples, i.e., forests with low naturalness, we consider the following sources: 
(1) \textit{Hyggen1990-2000}, forests harvested between 1990 and 2000;
(2) \textit{Pskog30till80}, forest stands between 30 and 80 years old; and
(3) \textit{BestandEjNaturvarden}, older forest stands that have been assessed with low natural value by forestry companies.
These areas represent a range of forests with different disturbance levels that compromise the naturalness value. 
Areas defined in \textit{Hyggen1990-2000} contain young forests, mostly coniferous-dominated, detected by a yearly change detection analysis from satellite images conducted by the Swedish Forestry Agency.

\subsection{Training-validation-test split}
\label{sec:tr-va-te-split}
We divided the data in training, validation, and test sets with a process based on geography.
First, we defined a uniform grid of $1.28 \times 1.28$~km squares in the study area.
Then, we discarded squares without any labeled polygon.
Finally, we randomly selected 64\% of these squares to build the training set, 16\% to the validation set, and 20\% to the test set.
In terms of size, the total surface areas are 
\SI{29409.28}{\kilo\meter\squared}, 
\SI{7353.14}{\kilo\meter\squared}, and 
\SI{9265.15}{\kilo\meter\squared}
for the training, validation, and test sets, respectively. These regions are shown in Figure~\ref{fig:study-area}.

\begin{table}[bt]
    \caption{Summary statistics of the training, validation, and test sets.\label{tab:datasets-stats}}
    \medskip
    \small
    \newcolumntype{L}{>{\arraybackslash}X}
    \begin{tabularx}{\textwidth}{Lrrr|rrr}
    \toprule
    \textbf{Split} & \multicolumn{3}{c}{\textbf{Number of polygons}} & \multicolumn{3}{c}{\textbf{Total area (\SI{}{\kilo\meter\squared})}} \\
    & Total & High nat. & Low nat. & Total\textsuperscript{*} & High nat. & Low nat. \\
    \midrule
    Training & 49,405 & 18,303 & 31,102 & 1,911 & 875 & 1,041 \\
    Validation & 13,239 & 5,020 & 8,219 & 496 & 239 & 258 \\
    Test & 16,516 & 6,416 & 10,100 & 592 & 276 & 317 \\
    \bottomrule
    \end{tabularx}
    \noindent{\footnotesize{\textsuperscript{*} The total area slightly differs from the sum of the high and low naturalness areas because of overlaps.}}
\end{table}

Labeled polygons have been divided into training, validation, and test sets depending on their geographical location.
Note that whenever a polygon belongs to two or more regions, we split the polygon accordingly.
So, for example, if part of a polygon lies in the training region and the other part lies in the validation region then we split it into two polygons: one for the training set and the other for the validation set.
Polygons with an area less than 
\SI{0.01}{\kilo\meter\squared} 
have been discarded because they are too small for their naturalness to be reliably assessed.
In addition, we discarded polygons where the CHM is not completely available and those with no trees, i.e., where all the values in the CHM are less than \SI{4}{m} (see Sect.~\ref{sec:featureExtraction} for an explanation of the choice of this value).

The main characteristics of the three sets thus obtained are listed in Table~\ref{tab:datasets-stats}.
Note that, since the data sources are not perfect and surveys have been made independently of each other, there are overlaps between polygons in the same set and even some (rare) areas that belong both to low and high naturalness polygons.
However, due to the splitting procedure, there are no overlaps among data sets. 
In other words, there are some geographical areas that belong to more than one polygon in the training set, for example, but there are no areas that belong both to a polygon in the training set and a polygon in the validation (or test) set.

\begin{figure}[h]
    \centering
    \resizebox{\linewidth}{!}{\subimport{fig/}{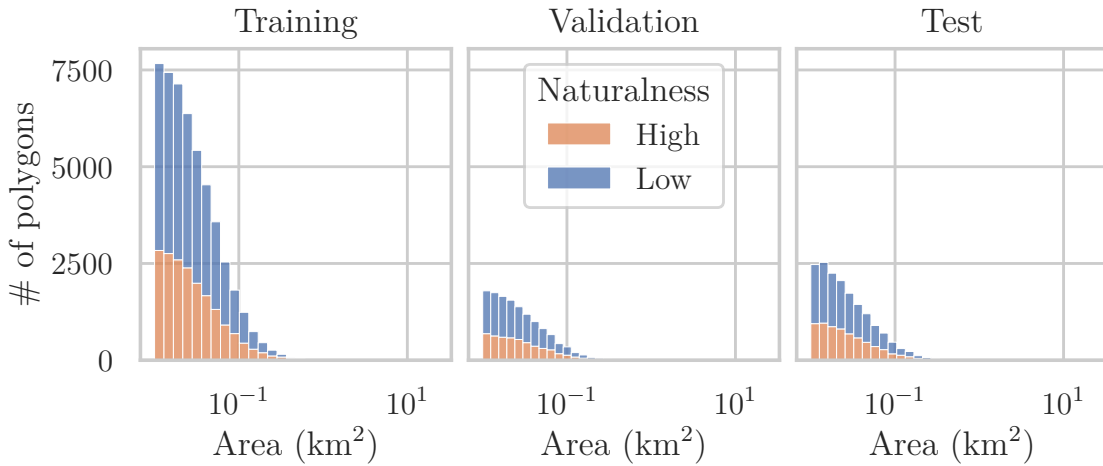}}
    \caption{Areas of the forests considered in training, validation, and test sets.}
    \label{fig:areas}
\end{figure}

Figure~\ref{fig:areas} shows the distribution of areas of the polygons in each split, overlaid by the naturalness (high or low) of the forest. 
As can be seen in the figure, the vast majority of areas are less than \SI{0.1}{\kilo\meter\squared} (ten hectares). 

\section{Method}
\label{sec:method}
The proposed method is based on the characterization of the forests using human-understandable features related to tree heights and the spatial distributions of trees.
In other words, given a region of the CHM defined by a polygon, we compute a set of values that are descriptive of the forest.
Then, from this set of values (i.e., the features) we predict the probability of the region to be a highly natural forest, by applying our classifier.
The classifier is trained using the dataset described in Section~\ref{sec:data}, containing positive and negative examples of high natural forests.

\begin{figure}[ht]
\resizebox{\linewidth}{!}{\subimport{fig/}{tree-density.pgf}}
    \caption{Example of tree density ($\rm{TD}=0.574$). The image on the left shows the CHM (shades of green) and the bounds of the region of interest (in red). The image on the right shows the area covered by trees (dark blue) and the remaining area (yellow) inside the region of interest, as well as the area outside the region of interest (light pink).
    \label{fig:tree-density}}
\end{figure}
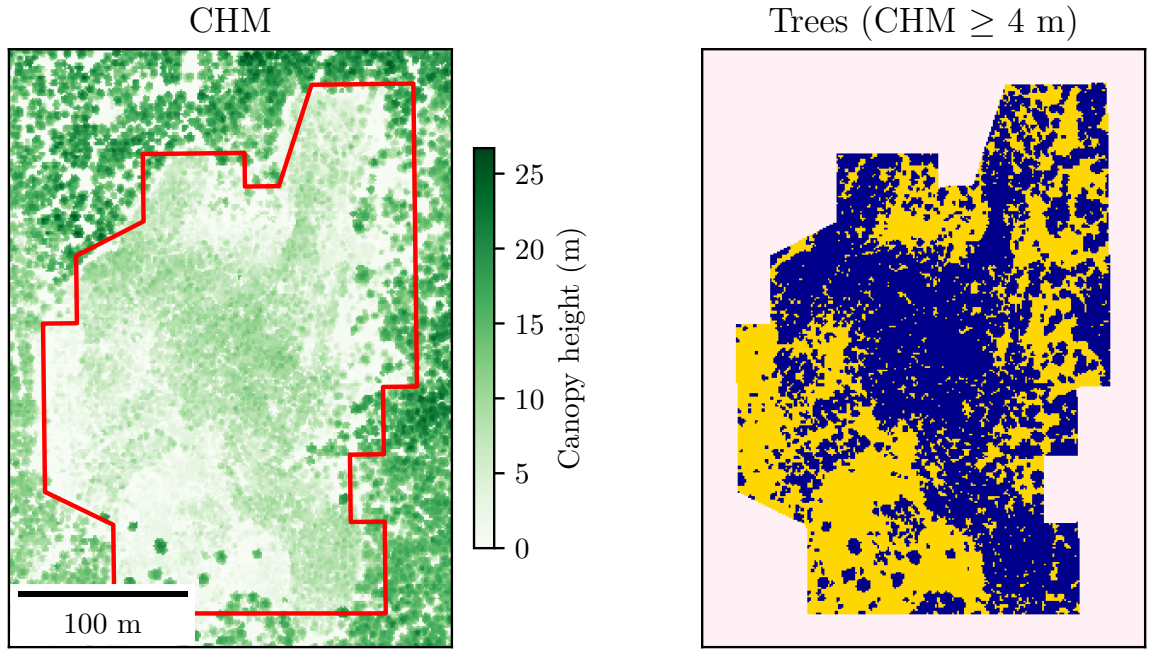

\begin{figure}[ht]
    \centering
    \resizebox{\linewidth}{!}{\subimport{fig/}{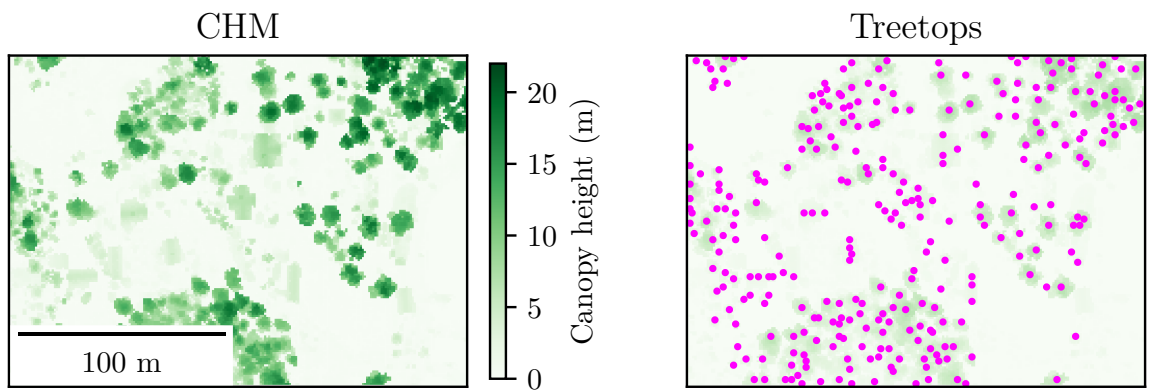}}
    \caption{Example of treetop identification. The image on the left shows the CHM. The image on the right shows the treetops (in magenta) identified in the same area.}
    \label{fig:treetops}
\end{figure}

\subsection{Feature extraction}
\label{sec:featureExtraction}
For any given polygon (see Section~\ref{sec:data}) several features are computed for the pixels in the CHM raster
that fall within the polygon, which defines the region of interest. In this process, we define a pixel as
a tree only if its value is equal to or greater than a threshold $h_{\rm{min}}$. 
Here, we have set this threshold as $h_{\rm{min}}=\,$\SI{4}{m}.
Note that for the type of forests in the study area, trees lower than \SI{4}{m} are very young.
See Figure~\ref{fig:tree-density} for an example of tree-labeled pixels.

In addition, we label some pixels as \emph{treetops}, so that we can distinguish single trees in the CHM.
Different methods have been proposed in the literature to identify single trees and treetops \citep{franceschiIdentifyingTreetopsAerial2018,ahmadiIndividualTreeDetection2022}.
Here, we identify treetops as a subset of the local maxima in the CHM raster. 
In particular, to compute the treetops, we apply a maximum filter with a window size of $w_s$ to the CHM image and we consider the pixels where all the following conditions are met: (i) value of the 
pixel in the filtered image is equal to the original; (ii) the value is equal to or greater than $h_{\rm{min}}$; (iii) the pixels are inside the region of interest.
Then, to avoid counting contiguous pixels with the same height as multiple treetops, we keep a single pixel for each region with contiguous maxima (considering 8-connectivity): the chosen pixel is the top-left-most.
Note that $w_s$ can be set as a parameter and it can be interpreted as $w_s = 2 d_{\rm{min}}+1$, where $d_{\rm{min}}$ corresponds to the minimal allowed distance separating treetops. Here, we have set $d_{\rm{min}}=\SI{2}{m}$.
This value has been chosen by visual inspection of the resulting treetops on some areas considering the CHM and orthophotos. Note that the resolution of the CHM is \SI{1}{m}, so it is not possible to select a lower value than \SI{1}{m} for $d_{\rm{min}}$.
See Figure~\ref{fig:treetops} for an example of treetop identification.
The features can then be computed, as will be described next.

\paragraph*{Tree density (TD)}
This feature measures the proportion of a polygon that is covered by trees. It is computed as
\begin{equation}
    \textrm{TD} = \frac{|H'|}{|H|},
\end{equation}
where $H$ is the set of values of the region of interest of the CHM, $H'$ is the set of values fulfilling the minimum height condition, i.e., 
\begin{equation}
\label{eq:Hmin}
    H'= \left\{ h \in H \; | \; h \geq h_{min} \right\},
\end{equation}
and the $|\cdot|$ notation denotes the cardinality of a set, i.e., the number of elements.
An example is shown in Figure~\ref{fig:tree-density}.

\paragraph*{Tree height mean (THM)}
This feature measures the average value of the CHM, considering only the pixels that represent trees, i.e., the average value over the elements of the set $H'$. Over the entire training set (all polygons) the mean of THM is equal to 12.4 m.

\paragraph*{Tree height variation (THV)}
As its name implies, this feature measures the variation in canopy height over the region of interest.
It is computed as the relative standard deviation (or coefficient of variation)
of the value of the pixels in the CHM, again considering only pixels 
with a value greater than $h_{min}$. So, given $H'$ as defined in Equation \ref{eq:Hmin}, we define 
\begin{equation}
    \textrm{THV} = \frac{\sigma(H')}{\textrm{THM}}
\end{equation}
where $\sigma(H')$ denotes the standard deviation over the elements of the set $H'$.

\paragraph*{Treetop density (TTD)}
The treetop density measures the density of individual trees.
It is computed from the treetop list, dividing the number of treetops inside the region of interest by its area:
\begin{equation}
    \textrm{TTD} = \frac{|T|}{A},
\end{equation}
where $T$ is the set of treetop heights inside the region of interest and $A$ is its area.
For this feature, we measure the area in hectares, as is standard practice in forestry.

\paragraph*{Treetop height mean (TTHM)}
This feature measures the average value of the treetop heights in the region of interest, i.e., the average value over the set $T$.
Over the entire training set (all polygons) the mean of TTHM is equal to 14.9 m.

\paragraph*{Treetop height variation (TTHV)}
This feature measures the variability of the treetop heights over the region of interest.
It is computed as the relative standard deviation of the treetop heights for trees that lie inside the region of interest, i.e., as
\begin{equation}
    \textrm{TTHV} = \frac{\sigma(T)}{\textrm{TTHM}}.
\end{equation}

\begin{figure}[ht]
    \centering
    \resizebox{\linewidth}{!}{\subimport{fig/}{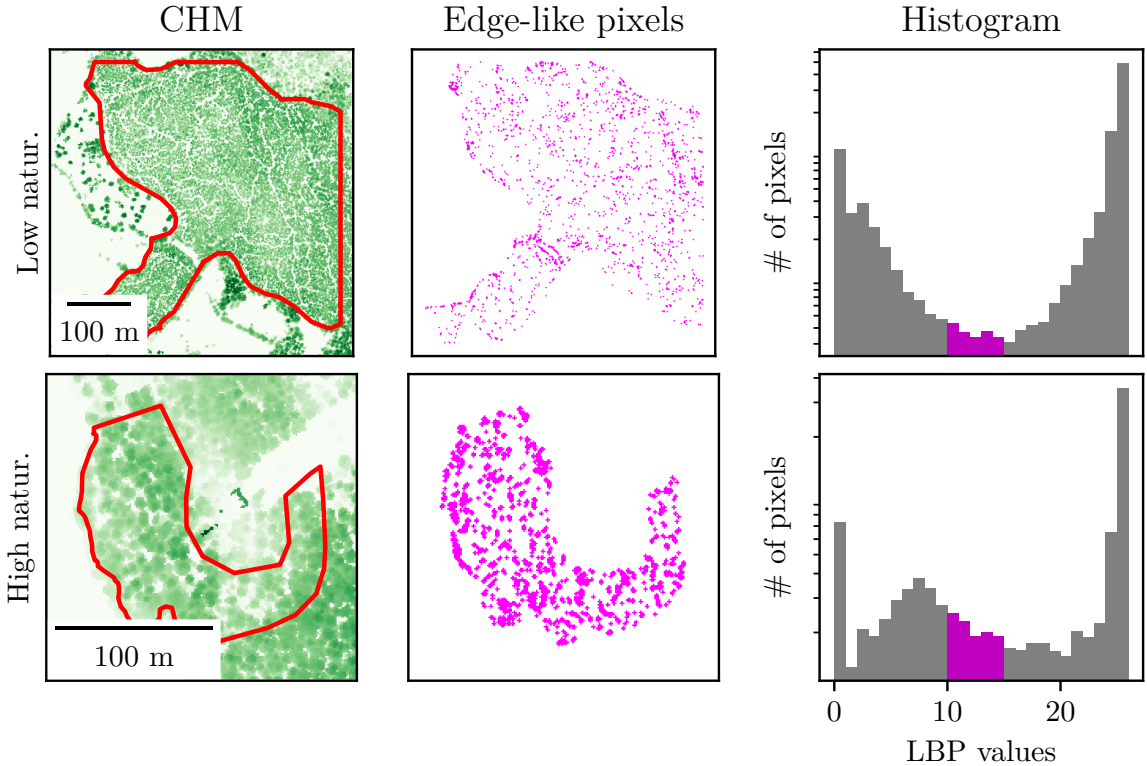}}
    \caption{Examples of the ELP feature for a low naturalness forest and a high naturalness one, shown in the top and in the bottom row, respectively. The edge-like pixels are shown in the middle column and their counts are highlighted in magenta in the histograms in the right column.}
    \label{fig:lbp}
\end{figure}

\paragraph*{Edge-like pixels (ELP)}
This feature considers the texture of the CHD and measures the proportion of pixels, within the region of interest, that are edge-like.
The intuition is that the human intervention in low naturalness forests results in more regular patterns in the CHD. 
So, the proportion of edge-like pixels is expected to be smaller in low naturalness forests because regular pattern has shorter edges than irregular patterns.
To identify edge-like pixels we use the local binary patterns (LBP) technique, which is a well-known method in computer vision~\citep{ojalaComparativeStudyTexture1996}.
In summary, the LBP method assigns to each pixel a value that depends on the difference with its neighbour pixels.
Two parameters can be set for defining the neighbor pixels, which lay in a circle centered in the considered pixel:
the radius $r$ of the circle
and the number $p$ of (circularly symmetric) points.
In our experiments we set $p=24$ and $r=3$.
The value of the ELP feature is then computed as the proportion of edge-like pixels, i.e. pixels with an LBP value in the range $p/2 \pm (r-1)$, which corresponds to the range [10, 14] with our choice of parameters.
Figure~\ref{fig:lbp} visualizes the LBP method for the computation of the ELP feature in two examples.

\begin{figure}[ht]
    \centering
    \footnotesize
    \includegraphics[width=\linewidth]{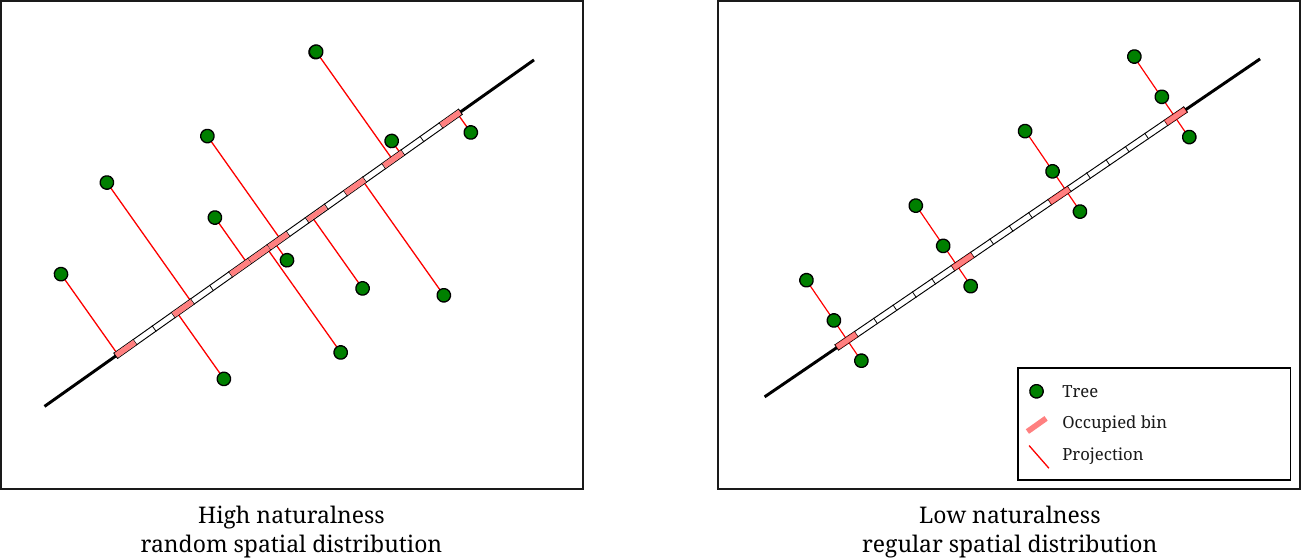}
    \caption{Visualization of the TTSD feature. On the left side treetops are distributed randomly and their projection on the line results in 9 of 18 bins being occupied. On the right side, the spacial distribution of the treetops is regular and it results in only 4 occupied bins.}
    \label{fig:ttsd}
\end{figure}

\paragraph*{Treetop spatial distribution (TTSD)}
This feature measures the regularity of the spatial distribution of treetops.
The intuition is similar to the previous feature: Human intervention results in a more regular spatial distribution of the trees w.r.t. natural grown forests. In order to measure the regularity, we project the treetop locations onto binned lines (in various directions $\alpha$) and we measure the fraction $f_w$ of bins that contain a (projected) treetop.
We consider a variable number of bins with a fixed width equal to $w$, in the range of treetop projections. 
We use $n$ lines with different directions $D_n$, equally spaced in the range $0-\ang{180}$.
The TTSD feature is then defined as:
\begin{equation}
\label{eq:TTSD}
    \textrm{TTSD} = \min_{\alpha \in D_n} f_w(\alpha)
\end{equation}
where the minimum $f_w$ is found at the direction that maximizes the regular spacing of the trees.
Note that the width $w$ of the bins and the number of lines $n$ can be seen as tunable parameters,
for which we have here used the values $w=\SI{1}{\meter}$ and $n=100$.
Figure~\ref{fig:ttsd} illustrates the intuition behind this feature.

\subsection{Machine learning models}
\label{sec:training}

Considering the annotated polygons from field inventories (see Section~\ref{sec:labels-data}) and their featurization as described above, we obtain a tabular data set in which forests (rows) are described by a set of 
features (columns) plus a binary label for naturalness, 
i.e., \textit{low} and \textit{high}, encoded as 0 and 1, respectively.

This results in a classification problem with tabular data, and many existing ML methods can be applied.
Since we are primarily interested in interpretable models, we considered perceptrons, logistic
regression, and decision trees.

\subsubsection{Perceptron}
\label{sect:perceptron}
A perceptron is a linear, binary classifier, in which a single hyperplane separates
the two classes~\citep{McCullochPitts1943}. Thus, the output from the perceptron
can be expressed as
\begin{equation} \label{eq:perceptron}
y = \Theta\left(\sum_{i=0}^{n}{w_i x_i}\right),
\end{equation}
where $\Theta$ denotes the Heaviside step function, $w_i$ are the tunable weights, $x_i, \,i = 1,\ldots n$ are the values of the $n$ features, and $x_0 \equiv 1$ (that multiplies the bias weight $w_0$), 
is introduced for convenience of notation.

The perceptron is trained using an iterative algorithm that is guaranteed to converge
if the two classes are fully separable. If
the classes are \textit{not} fully separable, as is normally the case in applied
problems, the algorithm does not formally converge but, provided that the number of
data points is finite, the weights will remain bounded.

The training algorithm starts by setting the weights to zero (or random, small values). 
Next, for each training epoch, the algorithm runs through all $m$ training examples,
after first placing them in random order.
For each training example $k$ ($k = 1,\ldots,m$) the weights are then updated as follows
\begin{equation}
w_i \leftarrow w_i + \eta\left(d_k - y_k\right)x_{i,k},\,\,i = 1,\ldots n
\end{equation}
where $d_k$ and $y_k$ are the desired and actual outputs (class labels, in this case) for
the $k^{\rm{th}}$ training example,
$\eta$ is the learning rate,
and $x_{i,k}$ are the feature values for the same
training example. 
The procedure is then repeated, epoch by epoch, until a termination
criterion is reached, for example a certain number of epochs.

\subsubsection{Logistic regression}
\label{sec:logregr}
Logistic regression is a linear classification method that is widely used in various fields. 
It is a statistical model that predicts the probability of a binary outcome, in this case high and low naturalness. 
The logistic regression model is characterized by its use of the logistic function, which is a sigmoid function that transforms a linear combination of input variables into a probability value between 0 and 1. 
This function is used to model the relationship between the input variables (the features) and the probability of the binary outcome, formally:
\begin{equation}
y = \Theta\left( \sigma(\beta^\intercal x) - 0.5 \right)
\end{equation}
where
$x$ is the input vector,
$\Theta$ denotes the Heaviside step function,
$\beta$ is the vector of coefficients, 
and $\sigma(t) = (1 + e^{-t})^{-1}$ is the logistic function.
Note that we rescale the feature values with a min-max normalization before applying the logistic regression, so that all components of $x$ have the same range.
In our experiments we use the scikit-learn\footnote{See \url{https://scikit-learn.org/stable/modules/generated/sklearn.linear_model.LogisticRegression.html} version 1.3.2.} default implementation.

\subsubsection{Decision trees}
Decision trees are a non-parametric machine learning method widely used for classification problems.
The model predicts the value of a target variable, the naturalness in our case, by learning simple decision rules inferred from the data features.
In our experiments we use the scikit-learn\footnote{See \url{https://scikit-learn.org/stable/modules/generated/sklearn.tree.DecisionTreeClassifier.html} version 1.3.2.} default implementation, which relies on the CART algorithm~\citep{cart1984}.

\subsection{Prediction of the probability of high naturalness}
\label{sec:confidenceScore}
The three ML models can provide a probability associated to the classification task, indicating the model's uncertainty about the predicted class.
Hence, the probability of being in the high naturalness class can be used to measure forest naturalness on a continuous scale, even if the training of the model considers only binary labels, i.e., positive and negative examples.
Moreover, this information can be valuable for decision-making, as it allows users to consider the uncertainty associated with the model's predictions.

In the case of logistic regression, the output of the logistic function is a value between zero and one, it can be interpreted as the probability $p_1$ of the observation to belong to Class 1 (high naturalness).
Note that, since our classification problem is a binary one, the probability of the observation to belong to Class 0 (low naturalness) is simply $p_0 = 1-p_1$.

In the case of the perceptron, the probability can be obtained by applying a sigmoid function (e.g., Gauss error function) to the results of the weighted sum (see Eq.~\ref{eq:perceptron}).

In the case of the decision trees, the probability can be obtained by limiting the depth of the tree and considering the fraction of training samples of the class in each leaf.

\subsection{Performance metrics}

We evaluate the proposed approach using holdout validation:
We use the training and validation set to build the models, and the test set for the evaluations.

We consider standard metrics commonly used for binary classification problems, namely
accuracy, precision, recall, and the F1 score.
Note that our data set is slightly unbalanced, since the low naturalness class represents more than half (about 62\% in all the splits) of the data, see Table~\ref{tab:datasets-stats}.
For this reason, we also considered the \textit{balanced accuracy}.

\section{Results}
\label{sec:results}

\subsection{Feature importance}
\label{sec:featureImportance}

We first analyse the discriminatory power of each feature, individually.
For each feature, we compute the \textit{optimal discrimination threshold} such that if we classify all the polygons below that threshold as low (or high) naturalness and vice-versa, the accuracy over the training set is maximized.
Results (see Table~\ref{tab:features}) show that features can be clearly divided into two categories: TTD, ELP, THM, and TTHM have a strong discriminatory power (accuracy in the range 84--85\%); while TTSD, TD, THV, and TTHV have less discriminatory power (accuracy less than 67\%).

Note that, compared to low-naturalness forests, those with high naturalness are characterized by larger values of tree and treetop height means, edge-like pixels, and randomness of treetop spatial distribution (THM, TTHM, ELP, TTSD).
By contrast, low-naturalness forests are characterized by larger values of treetop density (TTD).
High- and low-naturalness forests have similar values for tree and treeptop height variations, and tree density (THV, TTHV, TD).

\begin{table}[tb]
\caption{\label{tab:features}
Optimal discrimination threshold for each feature. The value of the threshold is shown in the third column. The threshold is computed considering the accuracy over training set (fourth column). The corresponding accuracy over the validation set is also shown (fifth column).
}
\medskip
\centering
\small
\newcolumntype{L}{>{\arraybackslash}X}
\begin{tabularx}{\textwidth}{Llrrr}
\toprule
Feature descr.& Feature & Threshold & Acc. (train.) & Acc. (valid.) \\
\midrule
Treetop density          & TTD & 214.2 & 0.8556 & 0.8569 \\
Edge-like pixels         & ELP & 0.061 & 0.8456 & 0.8473 \\
Tree height mean         & THM & 128.3 & 0.8454 & 0.8427 \\
Treetop height mean      & TTHM & 155.8 & 0.8452 & 0.8444 \\
\midrule
Treetop spatial distr.   & TTSD & 0.099 &  0.6658 & 0.6661 \\
Tree density             & TD & 0.165 & 0.6401 & 0.6338 \\ 
Treetop height variation & TTHV & 0.057 & 0.6305 & 0.6224 \\
Tree height variation    & THV & 0.752 & 0.6295 & 0.6208 \\
\bottomrule
\end{tabularx}
\end{table}

\begin{figure}[h!]
    \centering
    \resizebox{\linewidth}{!}{\subimport{fig/}{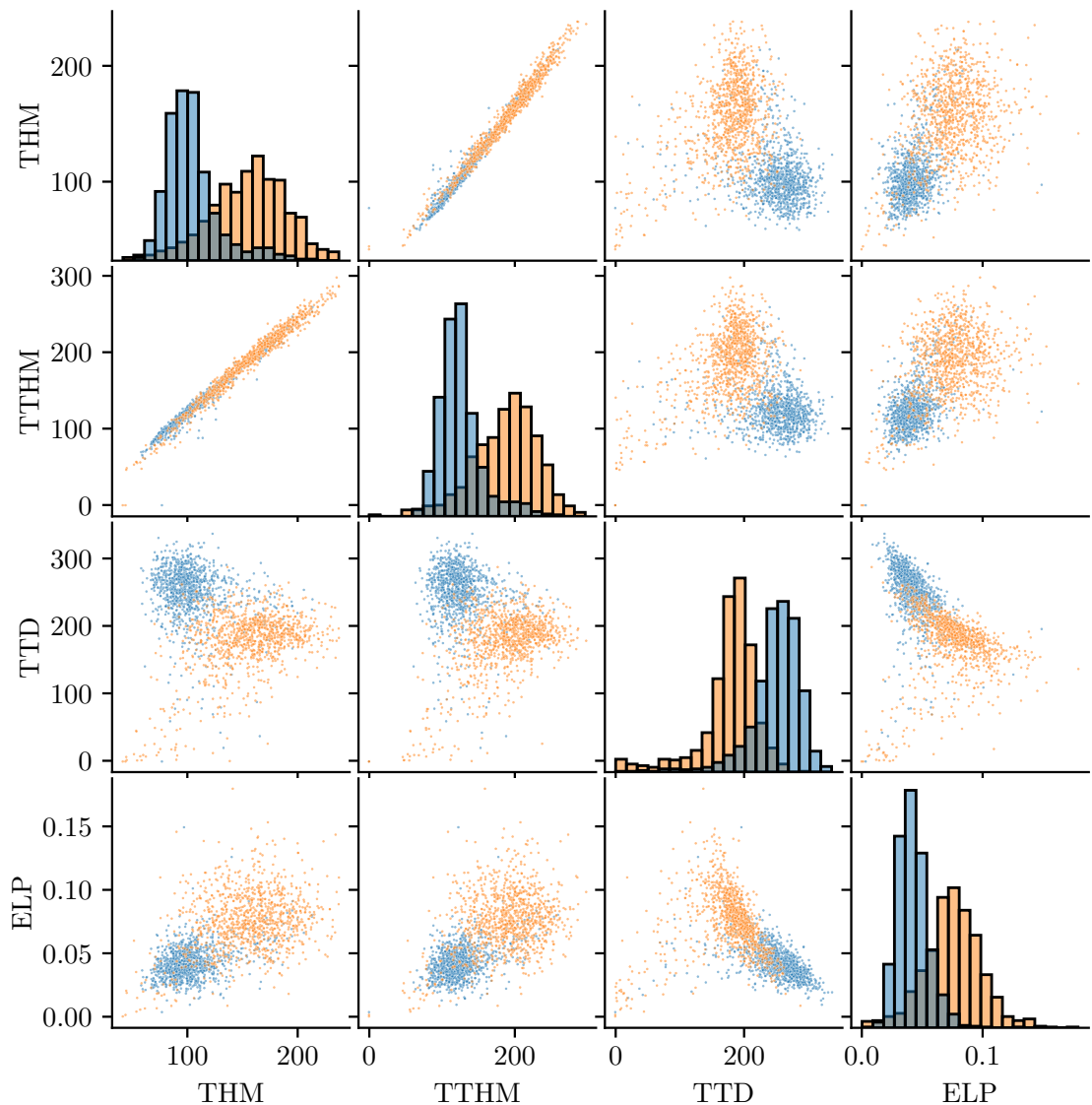}}
    \caption{Pair plot of the four most important features (see Table~\ref{tab:features}). Histograms for the individual features are shown on the diagonal, while the scatter plots for each pair of features are shown on the other charts. Orange points and bars represent high naturalness, while blue points and bars represent low naturalness.}
    \label{fig:pairplot}
\end{figure}

Figure~\ref{fig:pairplot} shows a pair plot of the four most important features, individually and in pairs.
Note that for visual clarity, a random balanced subset (2,000 data points) of the training set has been considered when generating the figure. It is interesting also to note
the good separation between the data points (for each class) for some of the pairs, 
especially in the case of the TTHM-TTD pair.

\subsection{Model comparison}

Overall, the three models have similar performances: about 0.9\% accuracy on the test set for all the considered models. In detail, the perceptron scores 0.897\%, logistic regression scores 0.898\%, and the decision trees score 0.901\%. 
In what follows, we investigate how the area of the polygon and the tree density affect the performance of the models.
The results obtained for the different metrics (accuracy, precision, recall, F1, and balanced
accuracy) are similar and follow the same pattern. Thus, for the sake of clarity, we present only those related to accuracy.
Note that all the accuracy scores reported in what follows refer to the test set.

\subsubsection{Accuracy vs. area of the polygon}
The full data set used here involves a set of forested areas that differ widely in size, ranging from 0.01 km$^2$ (1 hectare) to 22.6 km$^2$, even though the vast majority of areas are under 0.1 km$^2$ in size (cf. Fig.~\ref{fig:areas}). 
It could be expected that larger
areas would be easier to classify reliably than smaller ones,
just because they contain more data.
In order to investigate this hypothesis, the areas were sorted in increasing order, after which a given percentile range, based on area, was extracted from the training set. 
Next, the smallest and largest areas ($a_{\rm{min}}$ and $a_{\rm{max}}$) 
found in the extracted training subset were logged. Then, all areas falling in that
range were extracted from the validation and test sets. Thus, this procedure resulted in
three subsets, with the same range of areas.  Optimization runs where then carried out,
using the results over the extracted validation subset to determine when to stop the
runs, after which the results over the corresponding test subset were measured.
Those results are shown in Table~\ref{tab:accuracyVsArea}. As can be seen in the table,
the accuracy increases as the size of the areas grows, reaching a
value of 0.9497 for the largest 10\%.

\begin{table}[th]
\caption{\label{tab:accuracyVsArea}
Accuracy over the test set obtained when training over different subsets (percentile ranges based on area) of the training set, using all eight features. In each case, the results pertain to the model with highest validation
accuracy. The second column shows the area of the smallest polygon
($a_{\rm{min}}$) included in the training subset. The upper limit is 22.61 km$^2$, i.e., the maximum size, in all cases.}
\medskip
\centering
\small
\begin{tabular}{cc|ccc}
\toprule
Percentile range & $a_{\rm{min}}$ (km$^2$) & Perceptron &  Log. regr. & D.T. \\
\midrule
$\,\,\,0-100$  &  $0.010$ & 0.8970 & 0.8975 & 0.9009 \\
$10-100$ &  $0.012$ & 0.9015 & 0.9023 & 0.9031 \\
$20-100$ &  $0.014$ & 0.9042 & 0.9057 & 0.9117 \\
$30-100$ &  $0.017$ & 0.9086 & 0.9110 & 0.9207 \\
$40-100$ &  $0.020$ & 0.9137 & 0.9157 & 0.9186 \\
$50-100$ &  $0.024$ & 0.9215 & 0.9210 & 0.9265 \\
$60-100$ &  $0.029$ & 0.9237 & 0.9255 & 0.9356 \\
$70-100$ &  $0.037$ & 0.9321 & 0.9317 & 0.9372 \\
$80-100$ &  $0.050$ & 0.9405 & 0.9365 & 0.9418 \\
$90-100$ &  $0.077$ & 0.9497 & 0.9441 & 0.9441 \\
\bottomrule
\end{tabular}
\end{table}

\subsubsection{Accuracy vs. feature set}

For the sake of the interpretability of the system, we are interested in reducing the complexity of the model, while not losing too much in accuracy.
For this reason we report here the accuracy of the three machine learning models, considering subsets of the available features.
In particular, we consider the first six features, which are arguably the most easy to understand since they involve standard operations (average and coefficient of variation), and are easily visualizable, making them comprehensible to most observers, in our assessment. 
We then consider the four most important features in terms of discriminatory power (see Section~\ref{sec:featureImportance}), namely THM, TTD, TTHM, and ELP.
Finally, we consider only two features among those, discarding the less interpretable ELP as well as THM that is highly correlated with TTHM.
Results are show in Table~\ref{tab:accuracyVsFeatures}.

\begin{table}[tbh]
\caption{\label{tab:accuracyVsFeatures}
Accuracy over the test set considering selected subsets of the available features.
The second column shows the number ($m$) of features in the subset.}
\medskip
\centering
\small
\begin{tabular}{lc|ccc}
\toprule
Feature set & $m$ & Perceptron & Log. regr. & D.T. \\
\midrule
All                           & 8 & 0.8970 & 0.8975 & 0.9009 \\
TD, THM, THV, TTD, TTHM, TTHV & 6 & 0.8951 & 0.8969 & 0.8946 \\
THM, TTD, TTHM, ELP           & 4 & 0.8894 & 0.8893 & 0.8707 \\
TTD, TTHM                     & 2 & 0.8875 & 0.8850 & 0.8515 \\
\bottomrule
\end{tabular}
\end{table}

\subsubsection{Accuracy vs. prediction confidence}
The predictions of the proposed ML models are associated with a probability (see Section~\ref{sec:confidenceScore}).
The \textit{confidence score} of the classification is defined as the probability of predicted class.
Note that in binary classification the confidence score can take values from 0.5 to 1: values close to 0.5 when the outcome of the model is very uncertain; values close to 1 when there is no doubt.
We naturally expect the higher the confidence, the higher the accuracy.
Table~\ref{tab:accuracyVsConfidence} shows the accuracy of the complete (all eight features) logistic regression model as a function of the confidence score.
Note that the model is confident about the prediction most of the times: In fact, for about 67\% of the observations in the test set, the confidence score is greater than 0.9.
In these cases, the performance of the model is the highest, namely 96.7\%.
Notably the accuracy matches the range of the confidence score, as expected.

\begin{table}[tbh]
\caption{\label{tab:accuracyVsConfidence}
Accuracy over the test set considering the confidence of the prediction for the logistic regression model with all the features. The Support columns report the number of observations in absolute and relative terms.}
\medskip
\centering
\small
\begin{tabular}{crrr}
\toprule
Confidence score range & Support & Support \% & Accuracy \\
\midrule
0.5 $-$ 0.6 & 913 & 5.5\% & 0.5597 \\
0.6 $-$ 0.7 & 983 & 6.0\% & 0.6501 \\
0.7 $-$ 0.8 & 1,374 & 8.3\% & 0.7817 \\
0.8 $-$ 0.9 & 2,202 & 13.3\% & 0.8728 \\
0.9 $-$ 1.0 & 11,044 & 66.9\% & 0.9668 \\
\bottomrule
\end{tabular}
\end{table}

\section{Discussion}
\label{sec:discussion}

Our study presents an interpretable AI method for assessing the naturalness of forests from the CHM.
The method is composed of a feature extraction process and an ML model.
In our view, for high-stakes decision-making, a classifier should ideally be transparent and human-interpretable (as are our classifiers), in order to offer valuable insight, beyond merely a high probability of correct classification which is all that black box models can offer.

Our results show that the performance of the three considered ML models are similar: An accuracy of 0.89--0.90 over the test set.
What is really crucial in the proposed method is the feature extraction process.
This process involves selecting and transforming the raster input data into a set of values (i.e., the eight features) that is suitable for the ML models. 
In our study, we used a combination of features extracted from the CHM, considering a region of interest delimited by a labeled polygon. 
The features include metrics such as tree height, canopy cover, and density. 
We found particularly important the procedure that identifies the treetops, which is the basis for two of the features with the most discriminatory power, namely Treetop density (TTD) and Treetop height mean (TTHM).
This procedure exploits the high resolution of the CHM (i.e., \SI{1}{\meter}), for which a tree normally occupies more than one pixel.

It should be noted that with the proposed methodology we cannot claim that it is possible to assess the naturalness of every forest.
What our model really does is to distinguish between the classes (low or high naturalness) of the data source, as we have defined them.
Nevertheless, we have considered data sources for labeling that include a range of forests with different levels of human interference that can compromise the naturalness value.
So, we believe that our dataset is representative of most types of forest in the region of study, i.e., Southern Sweden.
To apply this method in other regions of the world with different types of forest, it will be necessary to re-train the model with new, meaningful examples of forests typical of those regions.

Regarding the interpretability of the system, it should be noted that the degree of interpretability is, to a great extent, a subjective one. 
In the case considered here, two factors are involved, namely the interpretability of the
individual features \textit{and} the interpretability of the model that uses them. 
Starting with the latter, and in contrast to black box models such as DNNs, the models presented here are fully transparent, generating their output as a linear combination of, or a set of rules based on, a small set of features. 
As for the features, the first six (TD, THM, THV, TTD, TTHM, and TTHV) involve standard operations (means and variances), and are easily visualizable, making them transparent to most observers, in our assessment.
The remaining two features, ELP and TTSD, are a bit more complex, yet within reach of interpretation for an observer with some knowledge of image interpretation.
Importantly, using only the six first features, one obtains a test set accuracy in the range 0.895 to 0.897 (depending on the model).
Moreover, by further simplifying the model and considering just two of those features (namely, TTD and TTHM), the performance decreases only slightly to an accuracy of 0.88 for the linear models, which corresponds to a 1\% loss (see Table~\ref{tab:accuracyVsFeatures}).

In situations where, say, a forest owner is prevented by a forest authority from exploiting a
particular area, it is a large advantage for every stakeholder to have a transparent model of 
the kind considered here.
In fact, one can identify the exact contribution of each feature to the final decision. 
Note that this is different from the post-hoc explanations methods for black box models, 
whose explanations are typically partial, often contradictory in cases where several explanation methods are considered~\citep{KrishnaEtAl2022}, and sometimes simply unreliable~\citep{SlackEtAl2020}.

Even though the explanation of a linear model is exact and readily understandable by a mathematically inclined observer, 
a lay person may require a \textit{verbal} explanation. Thus, in future work, we aim to combine our classifier with an 
equally transparent dialogue system (DAISY)~\citep{WahdeVirgolin2022}. This combination will result in a system that, 
when prompted, can provide a verbal explanation, an ability that is an integral part of the DAISY dialogue manager.

\section{Conclusion}
\label{sec:conclusion}

This paper presents an interpretable methodology for automated assessment of forest naturalness. 
The method uses eight features extracted from Canopy height model data and applies several different 
interpretable machine learning classifiers,
namely perceptrons, logistic regression, and decision trees.
The high-resolution data make it possible to identify individual trees and treetops, facilitating comprehensive feature extraction. 

We find that the accuracy of the method is largely independent of the specific machine learning model employed, but
is influenced by the procedures used to extract the features from the CHM (treetop identification, in particular), the feature selection, the area of the region of interest, and the prediction confidence score. 
In the study area encompassing coniferous and deciduous forests in Southern Sweden, the proposed method achieved an overall accuracy of approximately 90\%. 

The interpretability of the method is a key feature that enhances transparency and facilitates understanding by various stakeholders, including environmental experts, policy makers, forestry companies, and engineers involved in method development and maintenance. 
Moreover, the interpretable nature of our approach also allows for deeper comprehension of the factors contributing to a forest's naturalness classification, empowering stakeholders to make informed decisions related to forest management and conservation efforts.

\section*{Acknowledgments}

We thank our colleagues from Skogsstyrelsen, especially Alice Högström, Liselott Nilsson, and Johan Häggmark (Combitech AB), who provided insight and expertise that greatly assisted the research, although the authors take full responsibility for all interpretations and conclusions of this paper.

\section*{Disclosure statement}
The authors report there are no competing interests to declare.

\section*{Funding}
This research was partially supported by VINNOVA, project 2022-01702.

\section*{Data availability statement}
The Canopy Height Model is provided by Skogsstyrelsen and it is publicly available at \url{https://www.skogsstyrelsen.se/sjalvservice/karttjanster/skogliga-grunddata/}, \textit{tr\"adh\"ojd} layer.

\newpage
\bibliography{references_zotero, references_other}

\end{document}

%% file: fig/tree-density.pgf
%% Creator: Matplotlib, PGF backend
%%
%% To include the figure in your LaTeX document, write
%%   \input{<filename>.pgf}
%%
%% Make sure the required packages are loaded in your preamble
%%   \usepackage{pgf}
%%
%% Also ensure that all the required font packages are loaded; for instance,
%% the lmodern package is sometimes necessary when using math font.
%%   \usepackage{lmodern}
%%
%% Figures using additional raster images can only be included by \input if
%% they are in the same directory as the main LaTeX file. For loading figures
%% from other directories you can use the `import` package
%%   \usepackage{import}
%%
%% and then include the figures with
%%   \import{<path to file>}{<filename>.pgf}
%%
%% Matplotlib used the following preamble
%%   \usepackage{unicode-math}
%%   \usepackage{fontspec}
%%   \makeatletter\@ifpackageloaded{underscore}{}{\usepackage[strings]{underscore}}\makeatother
%%
\begingroup%
\makeatletter%
\begin{pgfpicture}%
\pgfpathrectangle{\pgfpointorigin}{\pgfqpoint{5.336660in}{3.047557in}}%
\pgfusepath{use as bounding box, clip}%
\begin{pgfscope}%
\pgfsetbuttcap%
\pgfsetmiterjoin%
\definecolor{currentfill}{rgb}{1.000000,1.000000,1.000000}%
\pgfsetfillcolor{currentfill}%
\pgfsetlinewidth{0.000000pt}%
\definecolor{currentstroke}{rgb}{1.000000,1.000000,1.000000}%
\pgfsetstrokecolor{currentstroke}%
\pgfsetdash{}{0pt}%
\pgfpathmoveto{\pgfqpoint{0.000000in}{0.000000in}}%
\pgfpathlineto{\pgfqpoint{5.336660in}{0.000000in}}%
\pgfpathlineto{\pgfqpoint{5.336660in}{3.047557in}}%
\pgfpathlineto{\pgfqpoint{0.000000in}{3.047557in}}%
\pgfpathlineto{\pgfqpoint{0.000000in}{0.000000in}}%
\pgfpathclose%
\pgfusepath{fill}%
\end{pgfscope}%
\begin{pgfscope}%
\pgfsetbuttcap%
\pgfsetmiterjoin%
\definecolor{currentfill}{rgb}{1.000000,1.000000,1.000000}%
\pgfsetfillcolor{currentfill}%
\pgfsetlinewidth{0.000000pt}%
\definecolor{currentstroke}{rgb}{0.000000,0.000000,0.000000}%
\pgfsetstrokecolor{currentstroke}%
\pgfsetstrokeopacity{0.000000}%
\pgfsetdash{}{0pt}%
\pgfpathmoveto{\pgfqpoint{0.010000in}{0.010000in}}%
\pgfpathlineto{\pgfqpoint{2.082725in}{0.010000in}}%
\pgfpathlineto{\pgfqpoint{2.082725in}{2.829224in}}%
\pgfpathlineto{\pgfqpoint{0.010000in}{2.829224in}}%
\pgfpathlineto{\pgfqpoint{0.010000in}{0.010000in}}%
\pgfpathclose%
\pgfusepath{fill}%
\end{pgfscope}%
\begin{pgfscope}%
\pgfpathrectangle{\pgfqpoint{0.010000in}{0.010000in}}{\pgfqpoint{2.072725in}{2.819224in}}%
\pgfusepath{clip}%
\pgfsys@transformshift{0.010000in}{0.010000in}%
\pgftext[left,bottom]{\includegraphics[interpolate=true,width=2.073333in,height=2.820000in]{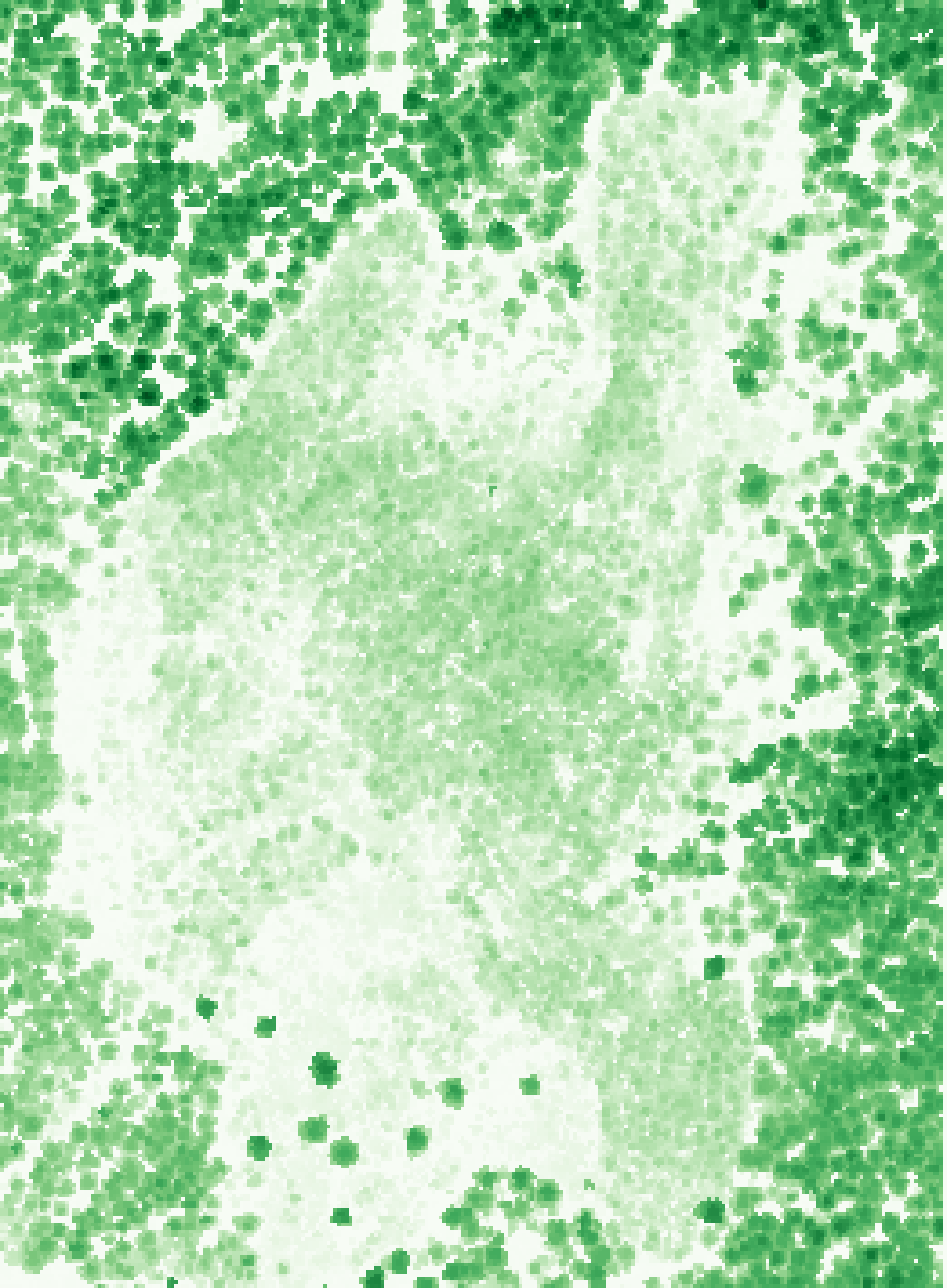}}%
\end{pgfscope}%
\begin{pgfscope}%
\pgfpathrectangle{\pgfqpoint{0.010000in}{0.010000in}}{\pgfqpoint{2.072725in}{2.819224in}}%
\pgfusepath{clip}%
\pgfsetrectcap%
\pgfsetroundjoin%
\pgfsetlinewidth{1.505625pt}%
\definecolor{currentstroke}{rgb}{1.000000,0.000000,0.000000}%
\pgfsetstrokecolor{currentstroke}%
\pgfsetdash{}{0pt}%
\pgfpathmoveto{\pgfqpoint{0.503228in}{0.168830in}}%
\pgfpathlineto{\pgfqpoint{0.498261in}{0.586724in}}%
\pgfpathlineto{\pgfqpoint{0.178899in}{0.741689in}}%
\pgfpathlineto{\pgfqpoint{0.169467in}{1.535377in}}%
\pgfpathlineto{\pgfqpoint{0.328205in}{1.537263in}}%
\pgfpathlineto{\pgfqpoint{0.324432in}{1.854739in}}%
\pgfpathlineto{\pgfqpoint{0.640020in}{2.017250in}}%
\pgfpathlineto{\pgfqpoint{0.636247in}{2.334725in}}%
\pgfpathlineto{\pgfqpoint{1.112460in}{2.340385in}}%
\pgfpathlineto{\pgfqpoint{1.114347in}{2.181647in}}%
\pgfpathlineto{\pgfqpoint{1.273085in}{2.183533in}}%
\pgfpathlineto{\pgfqpoint{1.426163in}{2.661633in}}%
\pgfpathlineto{\pgfqpoint{1.902376in}{2.667293in}}%
\pgfpathlineto{\pgfqpoint{1.919355in}{1.238653in}}%
\pgfpathlineto{\pgfqpoint{1.760617in}{1.236767in}}%
\pgfpathlineto{\pgfqpoint{1.764390in}{0.919291in}}%
\pgfpathlineto{\pgfqpoint{1.605652in}{0.917405in}}%
\pgfpathlineto{\pgfqpoint{1.609425in}{0.599929in}}%
\pgfpathlineto{\pgfqpoint{1.768163in}{0.601816in}}%
\pgfpathlineto{\pgfqpoint{1.773309in}{0.168830in}}%
\pgfpathlineto{\pgfqpoint{0.503228in}{0.168830in}}%
\pgfusepath{stroke}%
\end{pgfscope}%
\begin{pgfscope}%
\pgfsetrectcap%
\pgfsetmiterjoin%
\pgfsetlinewidth{0.803000pt}%
\definecolor{currentstroke}{rgb}{0.000000,0.000000,0.000000}%
\pgfsetstrokecolor{currentstroke}%
\pgfsetdash{}{0pt}%
\pgfpathmoveto{\pgfqpoint{0.010000in}{0.010000in}}%
\pgfpathlineto{\pgfqpoint{0.010000in}{2.829224in}}%
\pgfusepath{stroke}%
\end{pgfscope}%
\begin{pgfscope}%
\pgfsetrectcap%
\pgfsetmiterjoin%
\pgfsetlinewidth{0.803000pt}%
\definecolor{currentstroke}{rgb}{0.000000,0.000000,0.000000}%
\pgfsetstrokecolor{currentstroke}%
\pgfsetdash{}{0pt}%
\pgfpathmoveto{\pgfqpoint{2.082725in}{0.010000in}}%
\pgfpathlineto{\pgfqpoint{2.082725in}{2.829224in}}%
\pgfusepath{stroke}%
\end{pgfscope}%
\begin{pgfscope}%
\pgfsetrectcap%
\pgfsetmiterjoin%
\pgfsetlinewidth{0.803000pt}%
\definecolor{currentstroke}{rgb}{0.000000,0.000000,0.000000}%
\pgfsetstrokecolor{currentstroke}%
\pgfsetdash{}{0pt}%
\pgfpathmoveto{\pgfqpoint{0.010000in}{0.010000in}}%
\pgfpathlineto{\pgfqpoint{2.082725in}{0.010000in}}%
\pgfusepath{stroke}%
\end{pgfscope}%
\begin{pgfscope}%
\pgfsetrectcap%
\pgfsetmiterjoin%
\pgfsetlinewidth{0.803000pt}%
\definecolor{currentstroke}{rgb}{0.000000,0.000000,0.000000}%
\pgfsetstrokecolor{currentstroke}%
\pgfsetdash{}{0pt}%
\pgfpathmoveto{\pgfqpoint{0.010000in}{2.829224in}}%
\pgfpathlineto{\pgfqpoint{2.082725in}{2.829224in}}%
\pgfusepath{stroke}%
\end{pgfscope}%
\begin{pgfscope}%
\definecolor{textcolor}{rgb}{0.000000,0.000000,0.000000}%
\pgfsetstrokecolor{textcolor}%
\pgfsetfillcolor{textcolor}%
\pgftext[x=1.046363in,y=2.912557in,,base]{\color{textcolor}\rmfamily\fontsize{12.000000}{14.400000}\selectfont CHM}%
\end{pgfscope}%
\begin{pgfscope}%
\pgfsetbuttcap%
\pgfsetmiterjoin%
\definecolor{currentfill}{rgb}{1.000000,1.000000,1.000000}%
\pgfsetfillcolor{currentfill}%
\pgfsetlinewidth{1.003750pt}%
\definecolor{currentstroke}{rgb}{1.000000,1.000000,1.000000}%
\pgfsetstrokecolor{currentstroke}%
\pgfsetdash{}{0pt}%
\pgfpathmoveto{\pgfqpoint{0.023889in}{0.023889in}}%
\pgfpathlineto{\pgfqpoint{0.873592in}{0.023889in}}%
\pgfpathlineto{\pgfqpoint{0.873592in}{0.300414in}}%
\pgfpathlineto{\pgfqpoint{0.023889in}{0.300414in}}%
\pgfpathlineto{\pgfqpoint{0.023889in}{0.023889in}}%
\pgfpathclose%
\pgfusepath{stroke,fill}%
\end{pgfscope}%
\begin{pgfscope}%
\pgfsetbuttcap%
\pgfsetmiterjoin%
\definecolor{currentfill}{rgb}{0.000000,0.000000,0.000000}%
\pgfsetfillcolor{currentfill}%
\pgfsetlinewidth{0.000000pt}%
\definecolor{currentstroke}{rgb}{0.000000,0.000000,0.000000}%
\pgfsetstrokecolor{currentstroke}%
\pgfsetstrokeopacity{0.000000}%
\pgfsetdash{}{0pt}%
\pgfpathmoveto{\pgfqpoint{0.051667in}{0.244444in}}%
\pgfpathlineto{\pgfqpoint{0.845814in}{0.244444in}}%
\pgfpathlineto{\pgfqpoint{0.845814in}{0.272637in}}%
\pgfpathlineto{\pgfqpoint{0.051667in}{0.272637in}}%
\pgfpathlineto{\pgfqpoint{0.051667in}{0.244444in}}%
\pgfpathclose%
\pgfusepath{fill}%
\end{pgfscope}%
\begin{pgfscope}%
\definecolor{textcolor}{rgb}{0.000000,0.000000,0.000000}%
\pgfsetstrokecolor{textcolor}%
\pgfsetfillcolor{textcolor}%
\pgftext[x=0.263602in,y=0.078611in,left,base]{\color{textcolor}\rmfamily\fontsize{10.000000}{12.000000}\selectfont 100 m}%
\end{pgfscope}%
\begin{pgfscope}%
\pgfsetbuttcap%
\pgfsetmiterjoin%
\definecolor{currentfill}{rgb}{1.000000,1.000000,1.000000}%
\pgfsetfillcolor{currentfill}%
\pgfsetlinewidth{0.000000pt}%
\definecolor{currentstroke}{rgb}{0.000000,0.000000,0.000000}%
\pgfsetstrokecolor{currentstroke}%
\pgfsetstrokeopacity{0.000000}%
\pgfsetdash{}{0pt}%
\pgfpathmoveto{\pgfqpoint{3.253935in}{0.010000in}}%
\pgfpathlineto{\pgfqpoint{5.326660in}{0.010000in}}%
\pgfpathlineto{\pgfqpoint{5.326660in}{2.829224in}}%
\pgfpathlineto{\pgfqpoint{3.253935in}{2.829224in}}%
\pgfpathlineto{\pgfqpoint{3.253935in}{0.010000in}}%
\pgfpathclose%
\pgfusepath{fill}%
\end{pgfscope}%
\begin{pgfscope}%
\pgfpathrectangle{\pgfqpoint{3.253935in}{0.010000in}}{\pgfqpoint{2.072725in}{2.819224in}}%
\pgfusepath{clip}%
\pgfsys@transformshift{3.253935in}{0.010000in}%
\pgftext[left,bottom]{\includegraphics[interpolate=true,width=2.073333in,height=2.820000in]{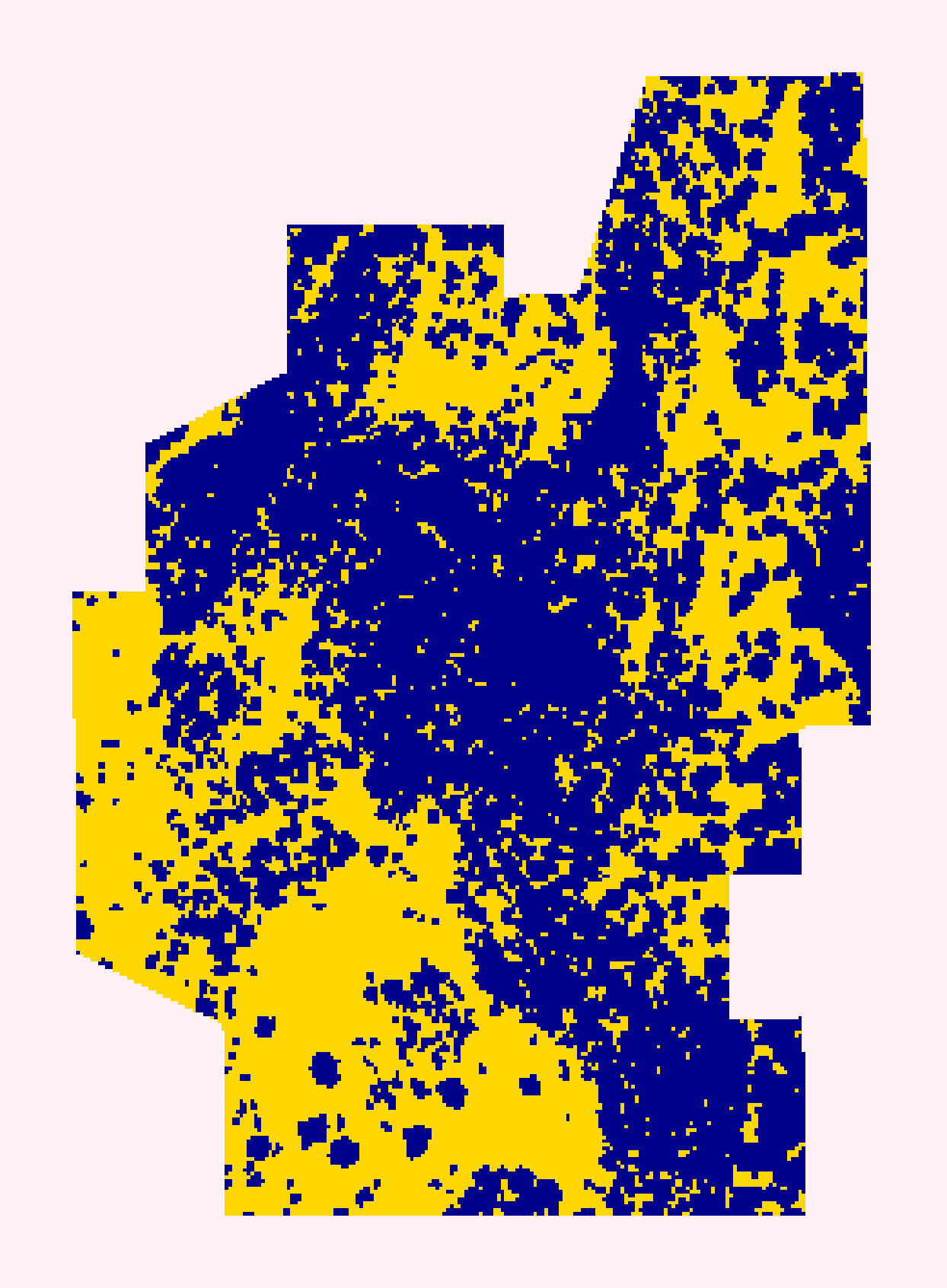}}%
\end{pgfscope}%
\begin{pgfscope}%
\pgfsetrectcap%
\pgfsetmiterjoin%
\pgfsetlinewidth{0.803000pt}%
\definecolor{currentstroke}{rgb}{0.000000,0.000000,0.000000}%
\pgfsetstrokecolor{currentstroke}%
\pgfsetdash{}{0pt}%
\pgfpathmoveto{\pgfqpoint{3.253935in}{0.010000in}}%
\pgfpathlineto{\pgfqpoint{3.253935in}{2.829224in}}%
\pgfusepath{stroke}%
\end{pgfscope}%
\begin{pgfscope}%
\pgfsetrectcap%
\pgfsetmiterjoin%
\pgfsetlinewidth{0.803000pt}%
\definecolor{currentstroke}{rgb}{0.000000,0.000000,0.000000}%
\pgfsetstrokecolor{currentstroke}%
\pgfsetdash{}{0pt}%
\pgfpathmoveto{\pgfqpoint{5.326660in}{0.010000in}}%
\pgfpathlineto{\pgfqpoint{5.326660in}{2.829224in}}%
\pgfusepath{stroke}%
\end{pgfscope}%
\begin{pgfscope}%
\pgfsetrectcap%
\pgfsetmiterjoin%
\pgfsetlinewidth{0.803000pt}%
\definecolor{currentstroke}{rgb}{0.000000,0.000000,0.000000}%
\pgfsetstrokecolor{currentstroke}%
\pgfsetdash{}{0pt}%
\pgfpathmoveto{\pgfqpoint{3.253935in}{0.010000in}}%
\pgfpathlineto{\pgfqpoint{5.326660in}{0.010000in}}%
\pgfusepath{stroke}%
\end{pgfscope}%
\begin{pgfscope}%
\pgfsetrectcap%
\pgfsetmiterjoin%
\pgfsetlinewidth{0.803000pt}%
\definecolor{currentstroke}{rgb}{0.000000,0.000000,0.000000}%
\pgfsetstrokecolor{currentstroke}%
\pgfsetdash{}{0pt}%
\pgfpathmoveto{\pgfqpoint{3.253935in}{2.829224in}}%
\pgfpathlineto{\pgfqpoint{5.326660in}{2.829224in}}%
\pgfusepath{stroke}%
\end{pgfscope}%
\begin{pgfscope}%
\definecolor{textcolor}{rgb}{0.000000,0.000000,0.000000}%
\pgfsetstrokecolor{textcolor}%
\pgfsetfillcolor{textcolor}%
\pgftext[x=4.290297in,y=2.912557in,,base]{\color{textcolor}\rmfamily\fontsize{12.000000}{14.400000}\selectfont Trees (CHM \(\displaystyle \geq\) 4 m)}%
\end{pgfscope}%
\begin{pgfscope}%
\pgfsetbuttcap%
\pgfsetmiterjoin%
\definecolor{currentfill}{rgb}{1.000000,1.000000,1.000000}%
\pgfsetfillcolor{currentfill}%
\pgfsetlinewidth{0.000000pt}%
\definecolor{currentstroke}{rgb}{0.000000,0.000000,0.000000}%
\pgfsetstrokecolor{currentstroke}%
\pgfsetstrokeopacity{0.000000}%
\pgfsetdash{}{0pt}%
\pgfpathmoveto{\pgfqpoint{2.186361in}{0.476609in}}%
\pgfpathlineto{\pgfqpoint{2.280662in}{0.476609in}}%
\pgfpathlineto{\pgfqpoint{2.280662in}{2.362614in}}%
\pgfpathlineto{\pgfqpoint{2.186361in}{2.362614in}}%
\pgfpathlineto{\pgfqpoint{2.186361in}{0.476609in}}%
\pgfpathclose%
\pgfusepath{fill}%
\end{pgfscope}%
\begin{pgfscope}%
\pgfsys@transformshift{2.186667in}{0.477557in}%
\pgftext[left,bottom]{\includegraphics[interpolate=true,width=0.093333in,height=1.886667in]{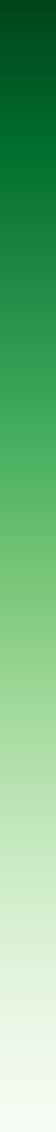}}%
\end{pgfscope}%
\begin{pgfscope}%
\pgfsetbuttcap%
\pgfsetroundjoin%
\definecolor{currentfill}{rgb}{0.000000,0.000000,0.000000}%
\pgfsetfillcolor{currentfill}%
\pgfsetlinewidth{0.803000pt}%
\definecolor{currentstroke}{rgb}{0.000000,0.000000,0.000000}%
\pgfsetstrokecolor{currentstroke}%
\pgfsetdash{}{0pt}%
\pgfsys@defobject{currentmarker}{\pgfqpoint{0.000000in}{0.000000in}}{\pgfqpoint{0.048611in}{0.000000in}}{%
\pgfpathmoveto{\pgfqpoint{0.000000in}{0.000000in}}%
\pgfpathlineto{\pgfqpoint{0.048611in}{0.000000in}}%
\pgfusepath{stroke,fill}%
}%
\begin{pgfscope}%
\pgfsys@transformshift{2.280662in}{0.476609in}%
\pgfsys@useobject{currentmarker}{}%
\end{pgfscope}%
\end{pgfscope}%
\begin{pgfscope}%
\definecolor{textcolor}{rgb}{0.000000,0.000000,0.000000}%
\pgfsetstrokecolor{textcolor}%
\pgfsetfillcolor{textcolor}%
\pgftext[x=2.377884in, y=0.428415in, left, base]{\color{textcolor}\rmfamily\fontsize{10.000000}{12.000000}\selectfont 0}%
\end{pgfscope}%
\begin{pgfscope}%
\pgfsetbuttcap%
\pgfsetroundjoin%
\definecolor{currentfill}{rgb}{0.000000,0.000000,0.000000}%
\pgfsetfillcolor{currentfill}%
\pgfsetlinewidth{0.803000pt}%
\definecolor{currentstroke}{rgb}{0.000000,0.000000,0.000000}%
\pgfsetstrokecolor{currentstroke}%
\pgfsetdash{}{0pt}%
\pgfsys@defobject{currentmarker}{\pgfqpoint{0.000000in}{0.000000in}}{\pgfqpoint{0.048611in}{0.000000in}}{%
\pgfpathmoveto{\pgfqpoint{0.000000in}{0.000000in}}%
\pgfpathlineto{\pgfqpoint{0.048611in}{0.000000in}}%
\pgfusepath{stroke,fill}%
}%
\begin{pgfscope}%
\pgfsys@transformshift{2.280662in}{0.829794in}%
\pgfsys@useobject{currentmarker}{}%
\end{pgfscope}%
\end{pgfscope}%
\begin{pgfscope}%
\definecolor{textcolor}{rgb}{0.000000,0.000000,0.000000}%
\pgfsetstrokecolor{textcolor}%
\pgfsetfillcolor{textcolor}%
\pgftext[x=2.377884in, y=0.781599in, left, base]{\color{textcolor}\rmfamily\fontsize{10.000000}{12.000000}\selectfont 5}%
\end{pgfscope}%
\begin{pgfscope}%
\pgfsetbuttcap%
\pgfsetroundjoin%
\definecolor{currentfill}{rgb}{0.000000,0.000000,0.000000}%
\pgfsetfillcolor{currentfill}%
\pgfsetlinewidth{0.803000pt}%
\definecolor{currentstroke}{rgb}{0.000000,0.000000,0.000000}%
\pgfsetstrokecolor{currentstroke}%
\pgfsetdash{}{0pt}%
\pgfsys@defobject{currentmarker}{\pgfqpoint{0.000000in}{0.000000in}}{\pgfqpoint{0.048611in}{0.000000in}}{%
\pgfpathmoveto{\pgfqpoint{0.000000in}{0.000000in}}%
\pgfpathlineto{\pgfqpoint{0.048611in}{0.000000in}}%
\pgfusepath{stroke,fill}%
}%
\begin{pgfscope}%
\pgfsys@transformshift{2.280662in}{1.182978in}%
\pgfsys@useobject{currentmarker}{}%
\end{pgfscope}%
\end{pgfscope}%
\begin{pgfscope}%
\definecolor{textcolor}{rgb}{0.000000,0.000000,0.000000}%
\pgfsetstrokecolor{textcolor}%
\pgfsetfillcolor{textcolor}%
\pgftext[x=2.377884in, y=1.134784in, left, base]{\color{textcolor}\rmfamily\fontsize{10.000000}{12.000000}\selectfont 10}%
\end{pgfscope}%
\begin{pgfscope}%
\pgfsetbuttcap%
\pgfsetroundjoin%
\definecolor{currentfill}{rgb}{0.000000,0.000000,0.000000}%
\pgfsetfillcolor{currentfill}%
\pgfsetlinewidth{0.803000pt}%
\definecolor{currentstroke}{rgb}{0.000000,0.000000,0.000000}%
\pgfsetstrokecolor{currentstroke}%
\pgfsetdash{}{0pt}%
\pgfsys@defobject{currentmarker}{\pgfqpoint{0.000000in}{0.000000in}}{\pgfqpoint{0.048611in}{0.000000in}}{%
\pgfpathmoveto{\pgfqpoint{0.000000in}{0.000000in}}%
\pgfpathlineto{\pgfqpoint{0.048611in}{0.000000in}}%
\pgfusepath{stroke,fill}%
}%
\begin{pgfscope}%
\pgfsys@transformshift{2.280662in}{1.536163in}%
\pgfsys@useobject{currentmarker}{}%
\end{pgfscope}%
\end{pgfscope}%
\begin{pgfscope}%
\definecolor{textcolor}{rgb}{0.000000,0.000000,0.000000}%
\pgfsetstrokecolor{textcolor}%
\pgfsetfillcolor{textcolor}%
\pgftext[x=2.377884in, y=1.487968in, left, base]{\color{textcolor}\rmfamily\fontsize{10.000000}{12.000000}\selectfont 15}%
\end{pgfscope}%
\begin{pgfscope}%
\pgfsetbuttcap%
\pgfsetroundjoin%
\definecolor{currentfill}{rgb}{0.000000,0.000000,0.000000}%
\pgfsetfillcolor{currentfill}%
\pgfsetlinewidth{0.803000pt}%
\definecolor{currentstroke}{rgb}{0.000000,0.000000,0.000000}%
\pgfsetstrokecolor{currentstroke}%
\pgfsetdash{}{0pt}%
\pgfsys@defobject{currentmarker}{\pgfqpoint{0.000000in}{0.000000in}}{\pgfqpoint{0.048611in}{0.000000in}}{%
\pgfpathmoveto{\pgfqpoint{0.000000in}{0.000000in}}%
\pgfpathlineto{\pgfqpoint{0.048611in}{0.000000in}}%
\pgfusepath{stroke,fill}%
}%
\begin{pgfscope}%
\pgfsys@transformshift{2.280662in}{1.889347in}%
\pgfsys@useobject{currentmarker}{}%
\end{pgfscope}%
\end{pgfscope}%
\begin{pgfscope}%
\definecolor{textcolor}{rgb}{0.000000,0.000000,0.000000}%
\pgfsetstrokecolor{textcolor}%
\pgfsetfillcolor{textcolor}%
\pgftext[x=2.377884in, y=1.841153in, left, base]{\color{textcolor}\rmfamily\fontsize{10.000000}{12.000000}\selectfont 20}%
\end{pgfscope}%
\begin{pgfscope}%
\pgfsetbuttcap%
\pgfsetroundjoin%
\definecolor{currentfill}{rgb}{0.000000,0.000000,0.000000}%
\pgfsetfillcolor{currentfill}%
\pgfsetlinewidth{0.803000pt}%
\definecolor{currentstroke}{rgb}{0.000000,0.000000,0.000000}%
\pgfsetstrokecolor{currentstroke}%
\pgfsetdash{}{0pt}%
\pgfsys@defobject{currentmarker}{\pgfqpoint{0.000000in}{0.000000in}}{\pgfqpoint{0.048611in}{0.000000in}}{%
\pgfpathmoveto{\pgfqpoint{0.000000in}{0.000000in}}%
\pgfpathlineto{\pgfqpoint{0.048611in}{0.000000in}}%
\pgfusepath{stroke,fill}%
}%
\begin{pgfscope}%
\pgfsys@transformshift{2.280662in}{2.242532in}%
\pgfsys@useobject{currentmarker}{}%
\end{pgfscope}%
\end{pgfscope}%
\begin{pgfscope}%
\definecolor{textcolor}{rgb}{0.000000,0.000000,0.000000}%
\pgfsetstrokecolor{textcolor}%
\pgfsetfillcolor{textcolor}%
\pgftext[x=2.377884in, y=2.194337in, left, base]{\color{textcolor}\rmfamily\fontsize{10.000000}{12.000000}\selectfont 25}%
\end{pgfscope}%
\begin{pgfscope}%
\definecolor{textcolor}{rgb}{0.000000,0.000000,0.000000}%
\pgfsetstrokecolor{textcolor}%
\pgfsetfillcolor{textcolor}%
\pgftext[x=2.572328in,y=1.419612in,,top,rotate=90.000000]{\color{textcolor}\rmfamily\fontsize{10.000000}{12.000000}\selectfont Canopy height (m)}%
\end{pgfscope}%
\begin{pgfscope}%
\pgfsetrectcap%
\pgfsetmiterjoin%
\pgfsetlinewidth{0.803000pt}%
\definecolor{currentstroke}{rgb}{0.000000,0.000000,0.000000}%
\pgfsetstrokecolor{currentstroke}%
\pgfsetdash{}{0pt}%
\pgfpathmoveto{\pgfqpoint{2.186361in}{0.476609in}}%
\pgfpathlineto{\pgfqpoint{2.233511in}{0.476609in}}%
\pgfpathlineto{\pgfqpoint{2.280662in}{0.476609in}}%
\pgfpathlineto{\pgfqpoint{2.280662in}{2.362614in}}%
\pgfpathlineto{\pgfqpoint{2.233511in}{2.362614in}}%
\pgfpathlineto{\pgfqpoint{2.186361in}{2.362614in}}%
\pgfpathlineto{\pgfqpoint{2.186361in}{0.476609in}}%
\pgfpathclose%
\pgfusepath{stroke}%
\end{pgfscope}%
\end{pgfpicture}%
\makeatother%
\endgroup%